\documentstyle[prl,aps,epsfig,floats]{revtex}
\title{\bf Violations of local realism with quNits up to $N=16$.}
\author{T. Durt$^1$, D. Kaszlikowski$^2$, M. \.Zukowski
$^{3}$.}
\address{$^1$ FUND, Free University of
Brussels, Pleinlaan 2, B-1050 Brussels, Belgium,$^2$ Instytut Fizyki Do\'swiadczalnej, Uniwersytet Gda\'nski, PL-80-952, Gda\'nsk, Poland,
$^3$ Instytut Fizyki Teoretycznej i Astrofizyki, Uniwersytet Gda\'nski, PL-80-952, Gda\'nsk, Poland.}
\begin{document}
\maketitle

{\bf Abstract}: Predictions for systems in entangled states cannot be described in
local realistic terms. However, after admixing some noise such a description is possible.
We show that for two qu$N$its (quantum systems described by $N$ dimensional Hilbert spaces)
in a maximally entangled state the minimal admixture of noise increases monotonically with
$N$. The results are a direct extension of those of Kaszlikowski et. al.,
Phys. Rev. Lett. {\bf 85}, 4418 (2000),
where results for $N\leq 9$ were presented. The extension up to $N=16$ is possible when one
defines for each $N$ a specially chosen set of observables. We also present results concerning
the critical detectors efficiency beyond which a valid test of local realism for entangled qu$N$its
is possible.


In early 1990's Peres and Gisin \cite{PERES} have shown, that if one
considers certain {\em dichotomic} observables applied to maximally
entangled state of two qu$N$its (particles described by an $N$-dimensional Hilbert space),
the violation of local realism, or
more precisely of the CHSH inequalities, survives
the limit of $N\rightarrow\infty$ and is maximal there.
However, for any dichotomic
quantum observables the CHSH inequalities give violations bounded by the
Tsirelson limit \cite{TSIRELSON}, i.e., it is limited by the
factor of $\sqrt{2}$.
Therefore, the question whether the violation of local realism
increases with growing $N$ was still left open.

It has been recently shown \cite{Kaszlikowski00} that
one indeed observes an increase with $N$ of
the discrepancy between
quantum and local realistic description of two maximally entangled qu$N$its
observed via unbiased multiport beamsplitters \cite{MULTIPEXP}. The results
presented in \cite{Kaszlikowski00} have been obtained via a numerical method
of linear optimisation and have been limited to $N=9$ \cite{inequalities}.

In the present paper we extend the
computations up to $N=16$. In the case of the method presented in \cite{Kaszlikowski00}
for $N\geq 10$ the computational time was prohibitively long. We avoid this problem here by
a careful choice of a fixed set of two pairs of observables for each $N$. As a result
one can avoid the time consuming
search
for optimal sets of
observables, which was a part of the computer program used in \cite{Kaszlikowski00}.

Another critical parameter for any Bell-type test is the threshold efficiency of the detector
to make it an unconditionally valid test of local realism. The efficiency a detector is usually
defined as its probability to fire when the quantum particle enters it. The procedure used in
\cite{Kaszlikowski00} can be easily adapted to handle also the question of inefficient detectors.
We report here the threshold values of efficiency for $N$ up to $16$. It decreases with $N$, however
the decrease is very slow.

Let us consider two
qu$N$it systems described by the mixed states in the form
\begin{eqnarray}
&\rho_N(F_{N})=F_{N}\rho_{noise}+
(1-F_{N})|\Psi_{max}^N\rangle\langle\Psi_{max}^N|,&
\label{werner}
\end{eqnarray}
where $|\Psi_{max}^N\rangle$ is a maximally entangled two qu$N$it state,
$\rho_{noise}=\frac{1}{N^2}\hat{I}$,
and the positive parameter $F_{N}\leq1$ determines the ``noise fraction"
within  the full state.
The threshold minimal $F_{N}^{tr}$,
for which the state $\rho_N(F_{N})$ allows a local realistic model,
will be our numerical value of the strength of violation of local realism by the quantum
state $|\Psi_{max}\rangle$.
The higher is $F_N^{tr}$ the higher is the minimum noise admixture required
to hide the nonclassicality of the quantum prediction.

To overcome the mentioned Tsirelson limit one has to use non-dichotomic observables.
Here, as in
the previous work, we limit ourselves to observables
defined by unbiased multiport beamsplitters.

{\it Unbiased} $2N$-port beamsplitters
\cite{ZEILINGER93}
are devices with the following property: if one photon enters into
any single input port (out of the $N$), its chances of exit are
equally split between all $N$ output ports. {The unbiased multiports are an
operational
realization of the concept of {\it mutually unbiased bases}, see
\cite{IVANOVIC}. Such bases are "as
different as possible" \cite{PERESBOOK}, i.e. fully complementary.} The
50-50 beamsplitter is the simplest member of the family.

One can always build an unbiased multiport with the distinguishing trait that
the elements of its unitary transition matrix, ${\bf U}^{N}$, are
{\it solely} powers of the $N$-th root of unity
$
\gamma_N=exp{(i2\pi/N)},
$
namely
$
{\bf U}^{N}_{ji}= \frac{1}{\sqrt{N}}\gamma_N^{(j-1)(i-1)}.
$
Devices
endowed with such a matrix were proposed to be called Bell
multiports \cite{MULT}.

\begin{figure}[htbp]
\begin{center}
\includegraphics[angle=0, width=12cm]{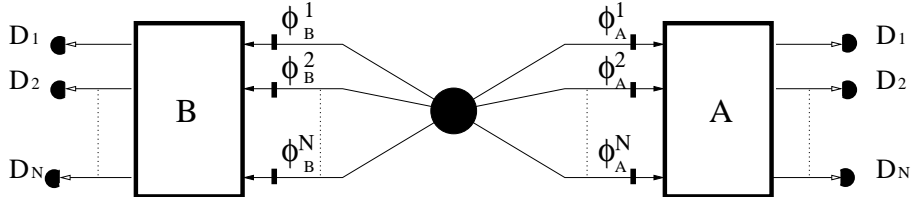}
\caption{The experiment of Alice and Bob with entangled qu$N$its.
Each of their measuring apparata consist of a set of $N$
phase shifters (PS) just in front of an $2N$ port Bell multiport, and
$N$ photon detectors $D_k, D_l$ (perfect, in the gedanken situation described
here) which register photons in the output ports of the device.
The phase shifters serve the role of the devices which
set the free macroscopic, classical parameters which can be
controlled by the experimenters. The source
produces a beam entangled two particle
state.}
\label{plotA}
\end{center}
\end{figure}

Let us now imagine spatially separated
Alice and Bob who
perform the experiment of (\ref{plotA}). The
maximally entangled state of the two qu$N$its

\begin{equation}
\mid\Psi^N_{max}\rangle={1\over\sqrt N}\sum^N_{m=1}\mid m,A\rangle\mid
m,B\rangle, \label{MULTISTATE}
\end{equation}

where e.g. $\mid m,A\rangle$ describes a photon in mode $m$ propagating to
Alice,
can be prepared with the aid of parametric down conversion (see
\cite{MULT}).
The two
sets of $N$ phase shifters at the inputs of the multiports, which
are denoted as $N$ dimensional ''vectors of phases"
$\vec{\phi}_A=(\phi^1_A,\phi^2_A,\dots,\phi^N_A)$ for Alice and
$\vec{\phi}_B=(\phi^1_B,\phi^2_B,\dots,\phi^N_B)$ for Bob,
introduce phase factor $e^{i(\phi^m_A+\phi^m_B)}$ in
front of the $m$-th component of the initial state (\ref{MULTISTATE}),
where $\phi^m_A$ and $\phi^m_B$ denote the local phase shifts. Alice
measures two observables $A_1, A_2$ defined by sets of phase shifts
$\vec{\phi}_{A_1}, \vec{\phi}_{A_2}$ whereas
Bob measures two observables $B_1, B_2$ defined by sets of the phase shifts
$\vec{\phi}_{B_1}, \vec{\phi}_{B_2}$.

Each set of local phase shifts constitutes the
interferometric realizations of the "knobs" at the disposal of
the observer controlling the local measuring apparatus which
incorporates also the Bell multiport and $N$ detectors.
In this way the local observable is defined. Its eigenvalues refer simply
to registration at one of the $N$ detectors behind the multiport. The
quantum prediction for the joint probability $P^{QM}_{F_N}(k,l)$ to
detect a photon at the $k$-th output of the multiport A and
another one at the $l$-th output of the multiport B calculated for the state (\ref{werner})
is given by:
\begin{eqnarray}
&P^{QM}_{F_N}{(k,l;\phi^1_A,...\phi^N_A,\phi^1_B,...\phi^N_B )}={1-F_N\over N}
\left|\sum^N_{m=1}\exp{[i(\phi^m_A+\phi^m_B)]} {\bf U}^N_{mk}{\bf
U}^N_{ml}\right|^2+{F_N\over N^2}&
\nonumber \\
& ={1-F_N\over N^3}\left(N+
2\sum^N_{m>n}\cos{({\bf \Phi}^m_{kl}-{\bf \Phi}^n_{kl})}\right)+{F_N\over N^2},&
\label{25a}
\end{eqnarray}
where
${\bf \Phi}^m_{kl} \equiv \phi^m_A+\phi^m_B +[m(k+l-2)] \frac{2\pi}{N}$.
The counts at a single detector, of course, do not depend upon
the local phase settings:
$
P^{QM}_{F_N}{(k)}=P^{QM}{(l)}_{F_N}={1}/{N}.$

The essential result of \cite{Kaszlikowski00} is that qu$N$its
violate local realism more strongly than qubits in the following sense:
the required minimal admixture of pure noise to the maximally entangled
state, such that a local realistic description of the quantum predictions becomes
possible, increases
with $N$. This result has been obtained via numerical methods of linear
optimisation. Here we give a brief account of the method
sending the reader for a more detailed description to
\cite{Kaszlikowski00}.

It is well known (see, e. g. \cite{FINE}, \cite{PERESBELL}) that
the hypothesis of local hidden variables
is equivalent to the existence of a (non-negative) joint
probability distribution involving all
four observables ($A_1,A_2,B_1,B_2$) from which it
should be possible to
obtain all the quantum predictions as marginals.
Let us denote this hypothetical joint distribution by
$P^{HV}(k_1,k_2,l_1,l_2)$, where $k_1$ and $k_2$
represent the outcome values for Alice's measurement of observables $A_1$ and $A_2$,
and $l_1$ and $l_2$
represents the outcome values for Bob's measurement of observables $B_1$ and $B_2$.
In quantum mechanics one cannot even define such a distribution, since
it involves mutually incompatible measurements.
A given set of quantum predictions, here $P^{QM}_{F_N}(k_i,l_j|A_i,B_j)$, is reproducible
by $P^{HV}(k_1,k_2,l_1,l_2)$, if and only if
\begin{eqnarray}
&P^{QM}_{F_N}(k_i,l_j|A_i,B_j)=\sum_{k_{i+1}}\sum_{l_{j+1}}P^{HV}(k_1,k_2,l_1,l_2),&
\label{sumation}
\end{eqnarray}
where $k_{i+1}$ and $l_{j+1}$ are understood as modulo $2$.
The Bell Theorem, within this context, says that there are quantum predictions,
which for $F_N$
below a certain threshold cannot be modelled by (\ref{sumation}), i.e.
there exists a critical $F_N^{tr}$ below  which one cannot have any local
realistic model. The $4N^2$
linear equations (\ref{sumation}) imposed on $N^4$ local hidden probabilities
$P^{HV}(k_1,k_2,l_1,l_2)$
form the full set of necessary and sufficient conditions
for the existence of local and realistic description of the
experiment.
This is a typical linear optimisation problem with $N^4+1$ non-negative unknowns,
$P^{HV}(k_1,k_2,l_1,l_2)$ and $F_N$,
and $4N^2$ linear conditions (\ref{sumation}).

In the previous work \cite{Kaszlikowski00} an involved computer algorithm \cite{GONDZIO1} was used 
to 
\begin{itemize}
\item (i) solve the linear optimization
problem for finding a minimal threshold $F_N^{tr}$ for which, under specific chosen settings,
(\ref{sumation}) is satisfied, 

\item (ii) find such settings for which (i) gives highest possible value
$F_N^{tr}$ (the so called ''amoeba" procedure was used \cite{Recipies}). 
\end{itemize}
Since the task (ii) makes the
computation, for high $N$, highly time consuming (since for each set of settings (i) has to be
solved), the results of \cite{Kaszlikowski00} reach only $N=9$.

Here we avoid this problem by dropping the point (ii) altogether. We search for $F_N^{tr}$
for a specific single set of observables $A^{(N)}_1,A^{(N)}_2,B^{(N)}_1,B^{(N)}_2$ for
each $N$.
We have used the phase settings in the following form: $\vec{\phi}_{A_1}=(0,0,\dots,0),
\vec{\phi}_{A_2}=(0,{\pi\over N},{2\pi\over N}\dots,{(N-1)\pi\over N})$ for Alice and
$\vec{\phi}_{B_1}=(0,{\pi\over 2N},{2\pi\over 2N},\dots,{(N-1)\pi\over 2N}),
\vec{\phi}_{B_2}=-\vec{\phi}_{B_1}$ for Bob. For $N=2$ 
$\phi_{A_1}^2=0,\phi_{A_2}^2={\pi\over 2}$,
$\phi_{B_1}^2={\pi\over 4},\phi_{B_2}^2=-{\pi\over 4}$. These are the standard phases for
the maximal violations of local realism in a two qubit experiment (the first phase in each 
''phase vector" is irrelevant). For $N=3$, $\vec{\phi}_{A_1}=(0,0,0)$,
$\vec{\phi}_{A_2}=(0,{\pi\over 3},{2\pi\over 3})$ and $\vec{\phi}_{B_1}=(0,{\pi\over 6},{\pi\over 3})$,
$\vec{\phi}_{B_2}=(0,-{\pi\over 6},-{\pi\over 3})$ give maximal violation of local realism (a result
of \cite{Kaszlikowski00} discussed in \cite{doktorat}). For $N\geq 4$ the phases were guessed.
However,
for up to $N=9$ these phases have
given exactly the same results as that obtained with the second stage of optimisation in
\cite{Kaszlikowski00}.
Of course, we do not know if they are really optimal for $N\geq 10$ because there is no
data for comparison. Nevertheless, the violation of local realism obtained for these
phases still grows with $N$ as it is depicted in FIG2. and the growth has the same
character as for $N\leq 9$.

\begin{figure}[htbp]
\begin{center}
\includegraphics[angle=0, width=11cm]{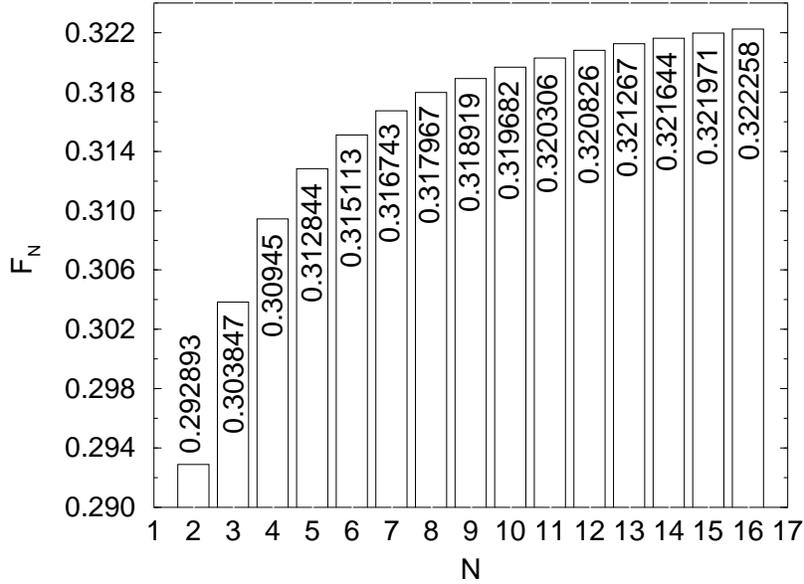}
\caption{Dependence of the critical noise admixture to the maximally entangled state 
(\ref{MULTISTATE}) on the dimension of the Hilbert space of a single subsystem. For larger noise
then shown here local realistic description exists. The increase of the value is interpreted
as an objective measure of the increasing with $N$ non-classicality of pairs of entangled qu$N$its.}
\label{plot0}
\end{center}
\end{figure}

Another interesting question that may be raised here concerns the critical quantum
efficiency of detectors below which there exists a local and realistic description
of the system. It was showed \cite{GargMermin} that for $N=2$ the critical efficiency
equals $2\sqrt2-2(\approx 0.828)$. Taking into account that violation of local realism grows
with $N$ one may expect that for higher dimensions of Hilbert space the critical
efficiency is lower than for two qubits. This problem has not been investigated
in our previous work. Here we show that the presented method can be just as well applied
to study this.

To this end it is necessary to modify
the conditions (\ref{sumation}) so as to take into account the probabilities
of non detection events, which are characterised by the quantum efficiency of
detectors $\eta$ ($0\leq\eta\leq 1$) (for simplicity we assume that the
efficiencies of all detectors are the same). This can be achieved as follows.
To a local non-detection event we ascribe the additional value that differs from the
values ascribed to the firings of detectors, say $0$. In this case there are more
local hidden probabilities and
more linear constraints imposed on them for now the indices enumerating possible
events extend from $0$ to $N$ (before the range was $1,\dots N$).


For non-ideal detectors, each endowed with identical inefficiency,
the quantum probabilities $P^{QM}_{F_N,\eta}(k_i,l_j|A_i,B_j)$ of coincidences between detector
$k_i$ at Alice's side and detector $l_j$ at Bob's side ($k_i,l_j\neq 0$)
while measuring observables $A_i,B_j$ are equal
to the corresponding probabilities with ideal detectors ($\eta=1$) multiplied by $\eta^2$, i.e.,
$P^{QM}_{F_N,\eta}(k_i,l_j|A_i,B_j)=\eta^2 P^{QM}_{F_N}(k_i,l_j|A_i,B_j)$. The quantum probabilities
$P^{QM}_{F_N,\eta}(0,l_j|A_i,B_j)$ and $P^{QM}_{F_N,\eta}(k_i,0|A_i,B_j)$ ($l_i\neq 0, k_i\neq 0$) of events
when one detector fails to fire at one of the sides of the experiment
equals
${1\over N}\eta(1-\eta)$ whereas the probability of the event when both detectors fail to fire
$P^{QM}_{F_N,\eta}(0,0|A_i,B_j)$
is $(1-\eta)^2$. Replacing left hand sides of (\ref{sumation}) by appropriate quantum
probabilities, i.e. $P^{HV}(k_i,l_j|A_i,B_j)=P^{QM}_{\eta,F_N}(k_i,l_j|A_i,B_j)$, one again obtains a
linear optimisation problem with respect to $F_N$, in which there are now $(N+1)^4$ local hidden
probabilities and $4(N+1)^2$ linear constrains.

Due to the fact that $\eta$ enters into equations quadratically it is not possible to
optimise it by means of linear programing methods.
The simple way of solving this difficulty is
the following. One decreases the value of $\eta$ (in our case by one percent) starting from $\eta=1$ and
keeping the local phases fixed until the program returns $F_N=1$, which signals that for this
efficiency there is already a local and realistic description.
Of course, the critical efficiency applies only to the case of the observables chosen here.
Once different observables or perhaps some non-maximally entangled state (compare \cite{Eberhard})
are chosen it may be lower.
The results are depicted in FIG3. We see that critical efficiency decreases very slowly but continuously
from the value obtained by Garg and Mermin \cite{GargMermin} for two qubits ($N=2$).

\begin{figure}[htbp]
\begin{center}
\includegraphics[angle=0, width=12cm]{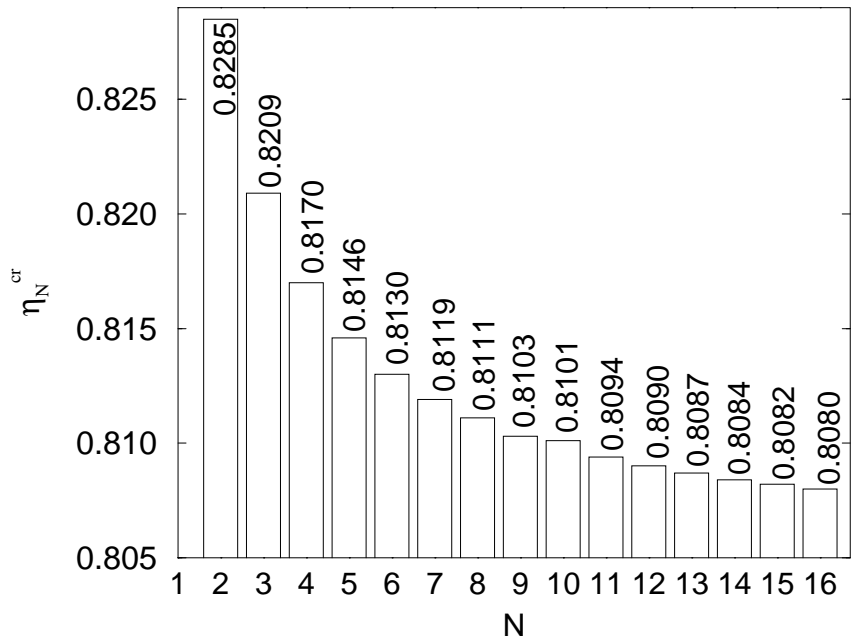}
\caption{Dependence of critical quantum efficiency of detectors $\eta_{N}^{cr}$ versus the dimension of the Hilbert 
space $N$.}
\label{plot1}
\end{center}
\end{figure}

{\it Acknowledgements}

MZ and DK are supported by the University of Gdansk Grant No
BW/5400-5-0032-0;  DK is supported by Fundacja na Rzecz Nauki
Polskiej.
TD is affiliated to the Fund
for Scientific Research (FWO), Flanders, as a post-doctoral fellow, and
member of the research group FUND (V.U.B.). This paper was written in the
framework of the Flemish-Polish
Scientific Collaboration Program No. 007.


\begin{references}
\bibitem{inequalities} The number of Bell
inequalities grows extremely
rapidly together with the increasing dimension $N$ of Hilbert space \cite{PERESBELL},
\cite{Pitovsky} and therefore their direct application becomes prohibitively
difficult
\end{references}
\end{document}